\documentclass[aps,10pt,pra,twocolumn,floatfix,floats,showpacs,superscriptaddress,raggedbottom]{revtex4-1}
\usepackage{graphicx,latexsym}
\usepackage{dcolumn}
\usepackage{amsmath}
\usepackage{amssymb,bm}
\usepackage{braket}
\usepackage{natbib}

\usepackage[hidelinks=true,breaklinks=true]{hyperref}
\hypersetup{
  colorlinks   = true, 
  urlcolor     = blue, 
  linkcolor    = blue, 
  citecolor    = red 
}

\usepackage[normalem]{ulem}

\def\sec#1{Sec.\ \ref{#1}}
\def\eq#1{Eq.\ (\ref{#1})}
\def\fig#1{Fig.\ \ref{#1}}

\begin{document}

\title{The interplay of electron-photon and cavity-environment coupling\\ on the 
       electron transport through a quantum dot system}

\author{Nzar Rauf Abdullah}
\email{nzar.r.abdullah@gmail.com}
\affiliation{Physics Department, College of Science, 
             University of Sulaimani, Kurdistan Region, Iraq}
\affiliation{Komar Research Center, Komar University of Science and Technology, Sulaimani, Iraq}

\author{Chi-Shung Tang}
\affiliation{Department of Mechanical Engineering,
  National United University, 1, Lienda, Miaoli 36003, Taiwan}

\author{Andrei Manolescu}
\affiliation{Reykjavik University, School of Science and Engineering,
              Menntavegur 1, IS-101 Reykjavik, Iceland}

\author{Vidar Gudmundsson}   
\email{vidar@hi.is}
 \affiliation{Science Institute, University of Iceland,
        Dunhaga 3, IS-107 Reykjavik, Iceland}
%

\begin{abstract}
We theoretically investigate the characteristics of the electron transport through a two-dimensional
a quantum dot system in the $xy$-plane coupled to a photon cavity and a photon reservoir, the environment.
The electron-photon coupling, $g_{\gamma}$, and the cavity-reservoir coupling, $\kappa$, are tuned to study 
the system in the weak, $g_{\gamma} \leq \kappa$, and the strong coupling regime, $g_{\gamma} > \kappa$. 
An enhancement of current is both seen with increasing $g_{\gamma}$ and $\kappa$ in the weak coupling regime 
for both $x$- and $y$-polarization of the photon field. This is a direct consequence of the Purcell effect. 
The current enhancement is due to the contribution of the photon replica states to the electron transport in which 
intraband transitions play an important role. 
The properties of the electron transport are drastically changed in the strong coupling regime with 
an $x$-polarized photon field in which the current is suppressed with increasing $g_{\gamma}$, but it is still increasing 
with $\kappa$. This behavior of the current is related to the population of purely electronic states and 
depopulation of photon replica states.
\end{abstract}



\maketitle

%
%

\section{Introduction}\label{Sec:Introduction}

The spontaneous emission of a quantum dots (QD) coupled to photon sources has been exploited 
to study the optical properties of QD materials \cite{Decker2013, Rahman_2001}. Controlling the spontaneous emission 
is a nontrivial task and is considered to be at the heart of quantum optics \cite{Lodahl2004} and  
is essential for diverse applications such as single-photon sources for quantum information \cite{PhysRevLett.89.067901, DeGreve2012, FriskKockum2019},
optical energy harvesting \cite{Graetzel2001}, and  light-emitting diodes \cite{Noda1123}.
The enhancement of a quantum system's spontaneous emission rate by its environment is called the Purcell effect \cite{PhysRev.69.37} and has been studied by many researchers in different systems
\cite{Kiraz_2003,PhysRevLett.120.114301}.
A strong Purcell enhancement is observed for quantum dots coupled to periodic metal-dielectric multi-layers that
can benefit various nanophotonic applications \cite{Li2019}.
The emission intensity of spin-dependent exciton states can be enhanced at a resonance 
due to the Purcell effect \cite{doi:10.1021/nl3008083}. 
In addition, the enhancement of the Purcell effect has been achieved 
by controlling the effective coupling with the micro-cavity \cite{PhysRevLett.98.063601,PhysRevLett.98.117402}.

The cavity quantum electrodynamics (QED) was proposed by Weisskopf and Wigner to control 
the spontaneous emission rate of a single photon source \cite{Weisskopf1930, Pelton2015}.
In cavity QED, the spontaneous emission rate is characterized by the mode density of the quantized 
photon field which is known as the Purcell effect in the context of cavity QED.
Recently, a QD system was used to enhance the spontaneous emission by a single photon nanocavity
coupled to the environment \cite{nano9050671}, where the coherency of the single photon source should be 
maintained \cite{Aspuru-Guzik2012, doi:10.1063/1.4977023}. 

High Purcell factor indicating a significant enhancement of an emission rate needs
a cavity confining light of small dimensions or storing light for a long time \cite{PhysRevB.94.235438}. 
This requires a high Q factor or a cavity with a nanoscale volume.
Generally, parameters should be carefully considered for obtaining a high Purcell factor,
such as the electron-photon coupling strength, $g_\gamma$, \cite{PhysRevB.91.205417}, and
the coupling strength of the cavity-photon field to the environment, $\kappa$ \cite{doi:10.1063/1.3294298,DELVALLE2011241}.
A system is said to be in the weak coupling regime if the electron-photon coupling strength is smaller than 
or equal to the coupling strength of the photon cavity to the environment, $g_{\gamma} \leq \kappa$, while 
in the strong coupling regime the condition should be $g_{\gamma} > \kappa$ \cite{PhysRevLett.121.043601}. 
In the weak coupling regime, the efficiency of a single photon
generation can be significantly enlarged, forming high-quality single photon generation by photon blockade
possible with current state-of-the-art samples \cite{PhysRevLett.114.233601}. Furthermore, 
the cavity-dot system can be made in a single step using a far-field optical lithography process monitoring 
a nonlinear optical transition of the QD system in which the observation of strong Purcell effect 
is evidenced \cite{PhysRevLett.101.267404}.
In contrast to the weak coupling regime, the Rabi oscillations in a QD exposed to a quantized photon field 
in the strong coupling regime lead to current peaks in connection with different types of photoluminescence 
\cite{Leng2018, PhysRevLett.116.113601, doi:10.1002/andp.201700334, PhysRevLett.109.240501, PhysRevLett.106.243601}. 
In addition, photon-assisted tunneling has been observed in a two-level system
with Rabi-effect \cite{Faraon2008} and also studied in a many level QD \cite{Vidar-ACS-Phot}.
Therefore, the strong coupling regime has been proposed for industrial technology \cite{Sillanpaa2007, Majer2007, Wu2019}.

The study of current transport in a QD-cavity is still lacking information
\cite{Hanschke2018, Lin2015} where the signs of Purcell effect appear. In this work, we consider a QD system 
coupled to a cavity and the photon reservoir in both the weak and the strong coupling regimes where the photon 
field is fully quantized \cite{Giannelli_2018, Kreinberg2018}. 
We compare the transport properties between these two regimes, where this comparison has not been 
demonstrated in our previous publications \cite{PhysicaE.64.254, GUDMUNDSSON_2019, nzar27.015301, Vidar85.075306, Abdullah2017}.
We assume a two-dimensional electron system consisting of a 
QD embedded in a short quantum wire and coupled to two electron reservoirs \cite{Zhang2016}. 
The transport properties of the QD system in the steady-state is investigated under 
the influence of a quantized photon field in a cavity using a Markovian quantum master equation~\cite{JONSSON201781}. 

We introduce the model system in~\sec{Sec:Model}.
Results are discussed for the model in~\sec{Sec:Results}. Finally, we have our conclusion in~\sec{Sec:Conclusion}.

\section{Theoretical formalism}\label{Sec:Model}

We consider a multi-level QD embedded in the center of a quantum wire in $xy$-plane
where the wire-dot system is hard-wall and parabolic confined in 
the $x$- and $y$-direction, respectively \cite{Vidar85.075306}.
The diameter of the QD is $d \backsimeq 66.5$~nm  and the length of the quantum wire 
is $L_x = 150$~nm.
The QD system is attached to two electron reservoirs, leads, from both ends in the $x$-direction. 
The total system, the QD and the leads, are microstructured in a GaAs heterostructure, and 
exposed to a perpendicular magnetic field in the $z$-direction, $B$.
The QD system is placed in a photon cavity that is coupled to a photon reservoir. 
The photon field is linearly polarized in the 3D rectangular cavity \cite{Abdullah2017,ABDULLAH2016280}. 
We let the photons in the cavity to be polarized in either $x$- or $y$-direction, of which
the latter is the direction of transport through the system.
The schematic diagram and the potential of the QD system coupled to the photon cavity can be seen 
in \cite{nano9050741}.

The Hamiltonian of the QD system coupled to the photon cavity is \cite{JONSSON201781, Vidar:ANDP201500298, 2040-8986-17-1-015201, GUDMUNDSSON20181672}  
\begin{align}
       H_\textrm{e} & =\sum_i \left(E_i+eV_{\mathrm{g}}\right)d_i^{\dagger}d_i  +
        \frac{1}{2}\sum_{ijrs}\langle V_{\textrm{Coul}}\rangle
        d_i^{\dagger}d_j^{\dagger}d_sd_r \, \nonumber \\
        & + H_Z + H_{\gamma} + g_\mathrm{{\gamma}}\sum_{ij}d_i^{\dagger}d_j\; g_{ij}
      \left\{a + a^\dagger\right\} \nonumber \\
      & + \frac{g_\mathrm{{\gamma}}^2}{\hbar\Omega_w} \sum_{i}d_i^{\dagger}d_i
      \left[  \hat{N}_{\gamma} + \frac{1}{2}\left( a^\dagger a^\dagger + aa  + 1 \right)\right],
    \label{H_total}
\end{align}
where the energy of a single-electron (SE) state and the gate voltage are presented as $E_i$, 
and $V_{\mathrm{g}}$, respectively. 
The Coulomb matrix elements in the SE state basis are
\begin{eqnarray}
      &&\langle V_{\mathrm{Coul}}\rangle = \langle ij|V_{\mathrm{Coul}}|rs\rangle   \nonumber \\
      &=& \int d\mathbf{r} d\mathbf{r^{\prime}} \psi^\mathrm{S}_{i} ({\bf r})^*
      \psi^\mathrm{S}_{j} ({\bf r}')^* V({\bf r} - {\bf r}')
      \psi^\mathrm{S}_r ({\bf r}') \psi^\mathrm{S}_s ({\bf r}),
\end{eqnarray}
with $\psi^\textrm{S}({\bf r})$ being the SE wavefunctions, the Coulomb interaction potential 
$V({\bf r} - {\bf r}')$, and $d_i^{\dagger}$($d_i$) being the electron creation (annihilation) operator 
of the QD system \cite{Vidar61.305}.
To treat the Coulomb interacting many-electron (ME) Hamiltonian in the QD system, an “exact numerical diagonalization” method is utilized \cite{Yannouleas70.2067}. 

In the second line of \eq{H_total}: $H_\textrm{Z} = g \mu_B B \sigma_z /2$ indicates the Zeeman Hamiltonian
where $g$ refers to the effective Land{\'e} g-factor, $\mu_B$ is the Bohr magneton, $B$ displays the weak 
external magnetic field, and $\sigma_z$ is a Pauli matrix. 
$H_{\gamma} = \hbar\omega_{\gamma} \hat{a}^{\dagger} \hat{a}$, 
introduces the photon field in the cavity where $\hbar\omega_{\gamma}$ indicates the photon energy and 
$\hat{a}^{\dagger}$ and $\hat{a}$ are the photon creation and annihilation operators, respectively.

The interaction between the electrons in the QD system and the photons in the cavity can be defined by the 
paramagnetic (last term of the second line of \eq{H_total} ) and the diamagnetic Hamiltonian (third line of \eq{H_total})
with $g_{ij}$ the dimensionless electron-photon coupling matrix, and $\hat{N}_{\gamma}$ the photon number operator~\cite{Vidar85.075306}.
The electron-photon coupling strength can be represented by $g_{\gamma}$ and $\Omega_w$ is the electron effective confinement frequency.

To investigate  the evolution of the electrons in the QD system in the steady state regime, we use a Markovian master equation where
the projection formalism is used based on the density operator \cite{Zwanzing.33.1338,Nakajima20.948}.
We are interested in the state of the central system, QD system after the coupling to the leads, 
the reduced density operator of the central system can be written as
\begin{equation}
 \hat{\rho}_\mathrm{S} = {\rm Tr_\mathrm{ L,R}}[\hat{\rho}(t)],
\end{equation}
where $\hat{\rho}(t)$ is the density operator of the total system. 
Here, the trace over the Fock space of the left (L) and the right (R) leads is taken into account \cite{GUDMUNDSSON20181672,PhysRev.129.2342,PhysRevA.31.3761,PhysRevA.84.043832},

The current from the left lead into the QD-system, $I^{c}_{\rm L}$,
and the current from it into the right lead, $I^{c}_{\rm R}$, can 
be introduced as 
\begin{equation}
 I^{c}_\mathrm{L,R} = {\rm Tr}_\mathrm{S} \Big( \Lambda^{\rm L,R}[\hat{\rho}_\mathrm{S};t] Q \Big),
\end{equation}
where $Q = -e \sum_i d_i^\dagger d_i$ is the charge operator of the wire-QD system 
with $\hat{d}^\dagger (\hat{d})$ the electron creation (annihilation) operator of the wire-QD system, respectively.  
The operators $\Lambda^\mathrm{L,R}$ represent the ``dissipation'' processes caused by both
leads~~\cite{GUDMUNDSSON_2019,JONSSON201781}.

\section{Results}\label{Sec:Results} 

We present the characteristics of the electron transport through the QD system in this section. 
We assume the system to be in a weak external perpendicular magnetic field, $B = 0.1$~T, 
to lift the spin degeneracy, but at the same time we want to avoid the effects of a strong Lorentz force 
on electron transport. 
The chemical potential of the left and the right leads are $\mu_L = 1.25$ and $\mu_R = 1.15$~meV, respectively, and
the temperature of the leads is fixed at $T_{L,R} = 0.5$~K. The gate voltage is $eV_{\rm g} = 0.615$~eV  that 
shifts the energy states of the QD system with respect to the chemical potential of the leads.
Furthermore, the photon energy is assumed to be $\hbar\omega_{\gamma} = 1.31$~meV. Under this condition 
the system is said to be in the off-resonant regime with respect to the lowest in energy electron states 
because the photon energy is smaller than their energy spacing.
The mean photon number in the photon reservoir is assumed to be $n_\mathrm{R} = 1$.

\subsection{Weak coupling regime, $g_{\gamma} \leq \kappa$}

\begin{figure}[htb]
\centering
    \includegraphics[width=0.25\textwidth]{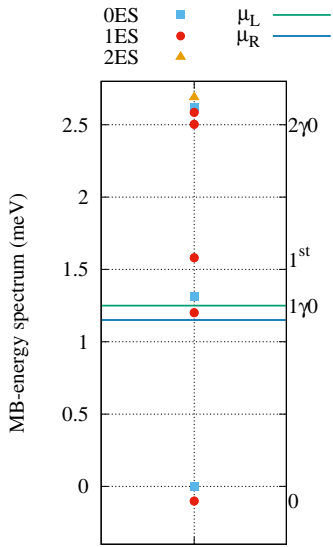}
 \caption{The many-Body energy spectrum of the closed QD system coupled to the 
  photon field, where 
  0ES (blue squares) are zero-electron states, 1ES (red circles) are one-electron state, 
  and 2ES are two-electron states (golden triangles). The photon energy is $\hbar \omega_{\gamma} = 1.31$~meV, $g_{\gamma} = 0.05$~meV, 
  $n_\mathrm{R} = 1$, and the photon field is polarized in the $x$-direction. 0 represents the one-electron
  ground-state, $1^{\rm st}$ is the one-electron first-excited state,
  and 1$\gamma$0 and 2$\gamma$0 refer to the first and second photon replicas
  of the one-electron ground-state, respectively.
  The chemical potential of the left lead is $\mu_L = 1.25$~meV (green line) and the right lead is $\mu_R = 1.15$~meV (blue line).
  The magnetic field is $B = 0.1~{\rm T}$, $eV_{\rm g} = 0.651$~meV, $T_{\rm L, R} = 0.5$~K, and $\hbar \Omega_0 = 2.0~{\rm meV}$.}
\label{fig01}
\end{figure}

In this section, we consider the electron-photon coupling strength, $g_{\gamma}$, to be smaller than 
or comparable to the selected range of the cavity-environment coupling strength, $\kappa$. 
The energy spectrum of the QD-cavity system as a function of  
$\kappa$ is shown in \fig{fig01}.
The cavity-environment coupling parameter, $\kappa$, is included in the Markovian master equation 
that is used to investigate the electron motion through the QD system here \cite{GUDMUNDSSON_2019}.
The energy spectrum of the closed system displayed in \fig{fig01} is independent of $\kappa$.
The green and the blue horizontal lines are the chemical potential of the left and the right leads, respectively.
0 indicates the one-electron ground-state, $1^{\rm st}$ labels the one-electron first-excited state,
and 1$\gamma$0 and 2$\gamma$0 are the first and the second photon replicas
of the one-electron ground-state, respectively. 
We should point out that each mentioned state includes both Zeeman spin components, spin-down and spin-up states, 
which are separated by the small Zeeman energy due to the presence of small external magnetic field.

Since the photon energy is smaller than the spacing between the lowest energy states of the 
QD system, $\hbar\omega_{\gamma} < E_j - E_i$, the one photon replica of the ground state, 1$\gamma$0, is located 
between 0 and $1^{\rm st}$, the system is in an off-resonant regime. 
Furthermore, the gate voltage, $eV_{g} = 0.615$~meV, moves up 
the 1$\gamma$0 into the  bias window, $eV_{\rm bias}= \mu_L -\mu_R$. 
As expected, the 1$\gamma$0 stays in the bias window when the cavity-environment coupling, $\kappa$, 
is varied, because changing $\kappa$ does not influence the physical characteristics of the closed QD system. 
We should remember that the energy spectrum for the $y$-polarized photon field is almost the same as 
for the $x$-polarizatized case for the low coupling regime, $g_{\gamma} \leq \kappa$, (not shown). 

\begin{figure}[htb]
\centering
    \includegraphics[width=0.4\textwidth,angle=0,bb=50 70 350 230]{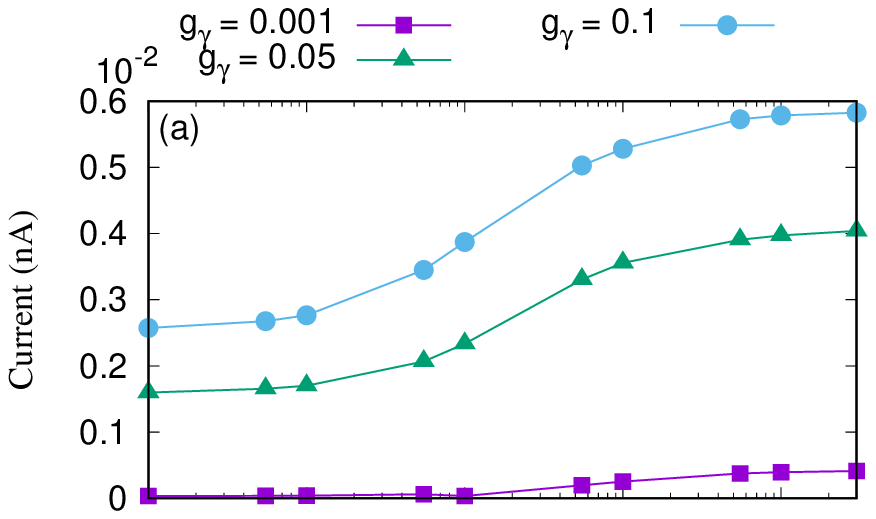}\\
    \includegraphics[width=0.4\textwidth,angle=0,bb=60 55 348 210]{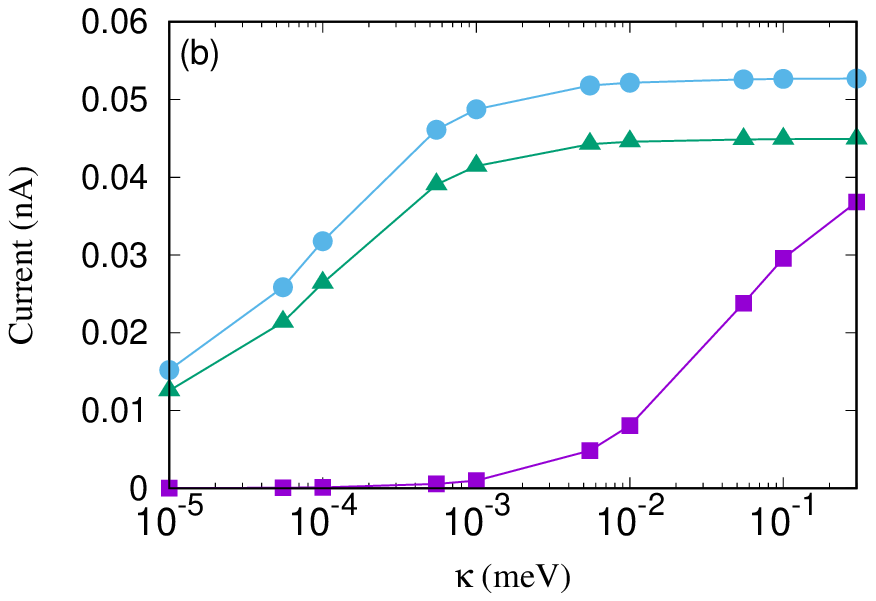}
 \caption{Current versus the cavity-environment coupling, $\kappa$,
          for $g_{\gamma} = 0.001$ (purple squares), $0.05$ (green triangles),  
          and $0.1$~meV (blue circles) in the $x$- (a) and $y$-polarized of the photon field. 
          The photon energy is $\hbar\omega_{\gamma} = 1.31$~meV, and $n_\mathrm{R} = 1$.
          The chemical potential of the left lead is $\mu_L = 1.25$~meV and the right lead is $\mu_R = 1.15$~meV.
          The magnetic field is $B = 0.1~{\rm T}$, $eV_{\rm g} = 0.651$~meV, $T_{\rm L, R} = 0.5$~K, and $\hbar \Omega_0 = 2.0~{\rm meV}$}
\label{fig02}
\end{figure}

The transport properties of the QD system are presented in \fig{fig02} in which the steady-state current 
as a function of the cavity-environment coupling is plotted for an $x$- (a) and $y$-polarized (b) photon field.
The current increases with $g_{\gamma}$ in the selected range of the cavity-environment coupling strength.
It can be clearly seen that the current in the $x$-polarized photon field is ten times larger than for the 
$y$-polarization for $g_{\gamma} = 0.05$ and $0.1$~meV. This is due to the fact that the states of the QD system are more polarizable in the $x$-direction and the photon replica states are actively contributing to the current 
transport compared to the $y$-polarization which will be shown later.
But for the very low coupling when $g_{\gamma} = 0.001$~meV, the current is almost the same for both  
photon polarizations since the photon field, and in turn the photon replica states, do not play an 
important role in the current transport for such a low $g_{\gamma}$.

\begin{figure}[htb]
\centering
    \includegraphics[width=0.22\textwidth]{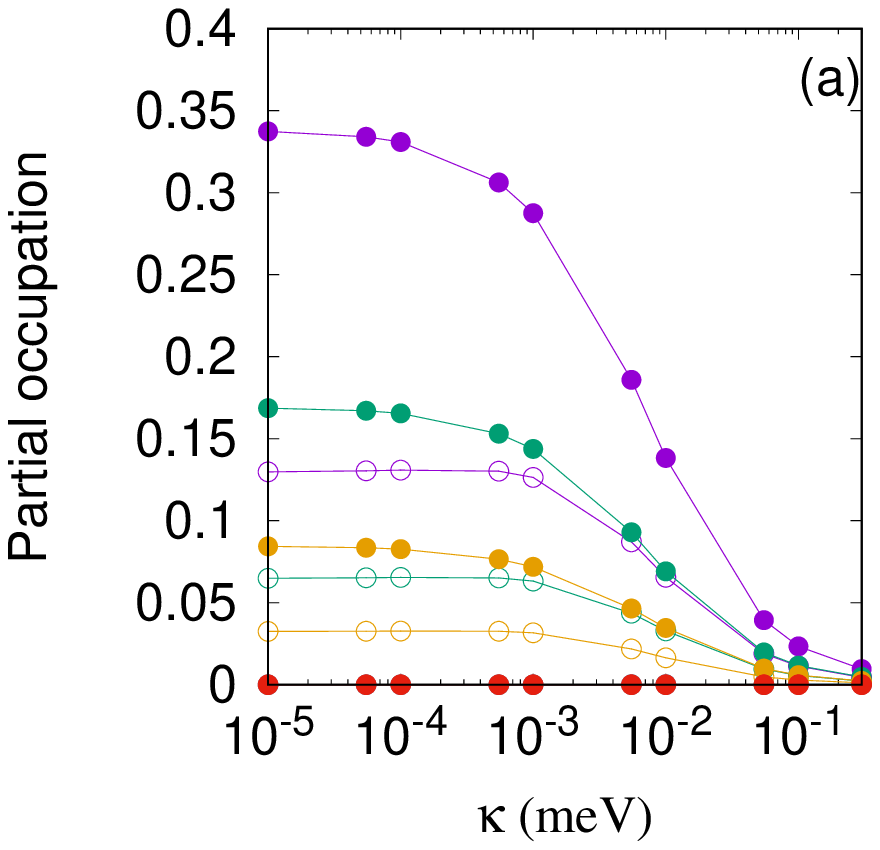}
    \includegraphics[width=0.24\textwidth]{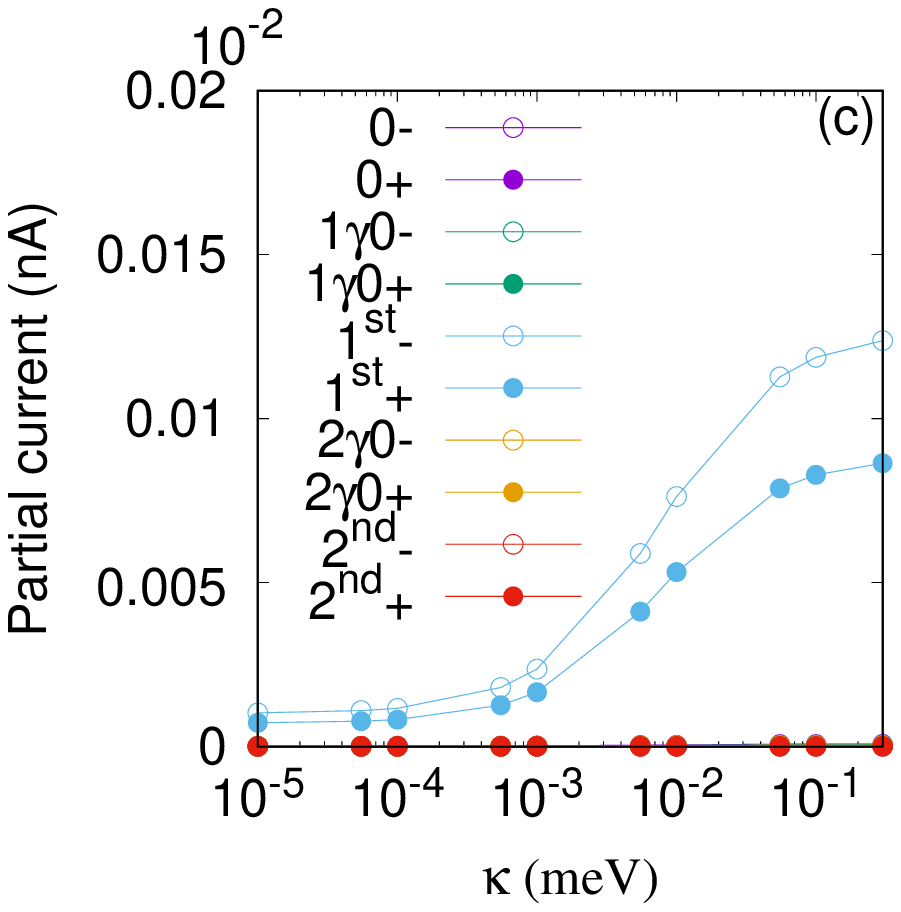}\\
    \includegraphics[width=0.22\textwidth]{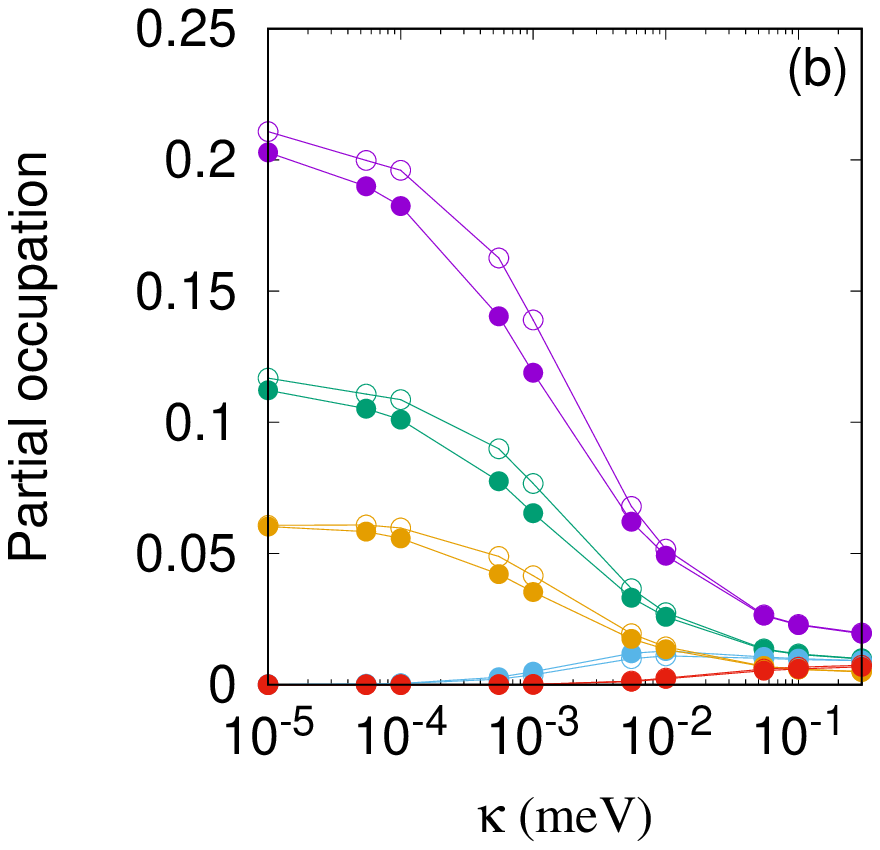}
    \includegraphics[width=0.23\textwidth]{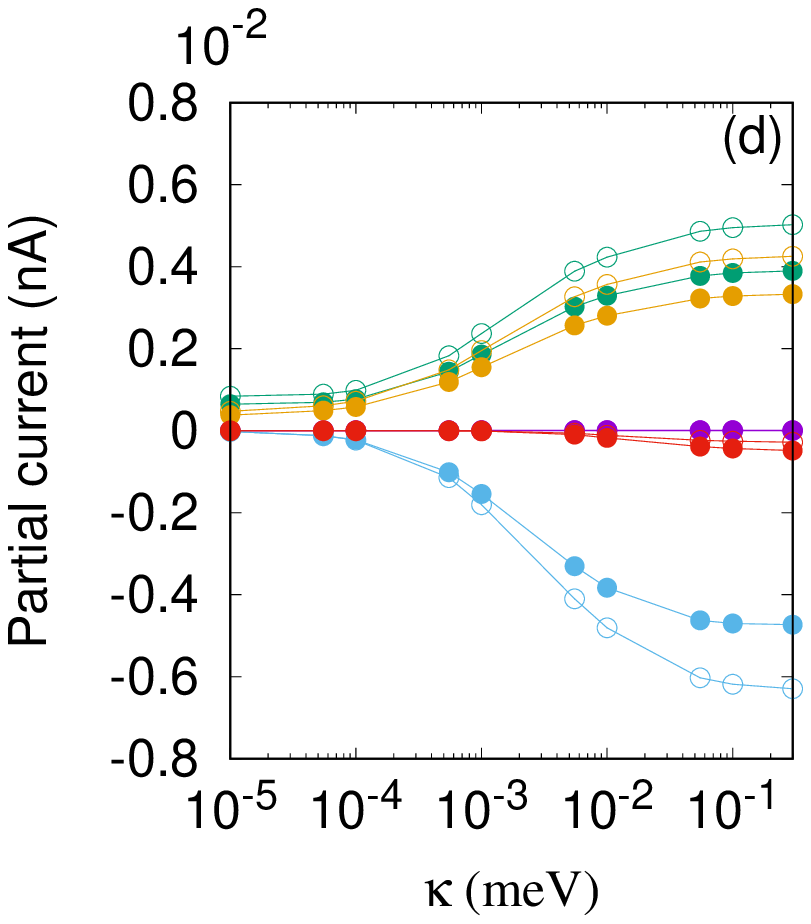}
 \caption{Partial occupation (a,b) and the partial current (c,d) of the ten 
          lowest one-electron states versus $\kappa$ are plotted 
          for $g_{\gamma} = 0.001$ meV (top panel) and $0.1$~meV (lower panel).
          The minus (-) and plus (+) signs indicate the Zeeman spin-down and up, respectively. 
          Herein, 0 displays the one-electron ground-state (purple circles) 
          $1\gamma$0 refers to the one-photon replica of the $0$  (green circles),
          $1^{\rm st}$ demonstrates the one-electron first-excited state (blue circles), 
          $2\gamma$0 is the two-photon replica of $0$ (golden circles),  
          and $2^{\rm nd}$ is the one-electron second-excited state (red circles).
          The photon energy is $\hbar\omega_{\gamma} = 1.31$~meV, $n_\mathrm{R} = 1$, and the photon field
          is polarized in the $x$-direction.
          The chemical potential of the left lead is $\mu_L = 1.25$~meV and the right lead is $\mu_R = 1.15$~meV.
          The magnetic field is $B = 0.1~{\rm T}$, $eV_{\rm g} = 0.651$~meV, $T_{\rm L, R} = 0.5$~K, and
          $\hbar \Omega_0 = 2.0~{\rm meV}$.}
\label{fig03}
\end{figure}

The most interesting point here is that the current is increased with the cavity-environment coupling which is a direct consequence 
of the Purcell effect. To better understand it the partial occupation (a,b) and the partial current (c,d) of the ten 
lowest one-electron states are plotted for $g_{\gamma} = 0.001$ (top panel) and $0.1$~meV (lower panel) in \fig{fig03} 
where the minus and the plus signs indicate the Zeeman spin-down and spin-up states, respectively. 
Let's first take a look the low coupling strength of the electron-photon, $g_{\gamma} = 0.001$~meV (\fig{fig03} (top panel).
Increasing $\kappa$, the occupation of all selected states is decreased as is shown in 
\fig{fig03}(a). It demonstrates that the lifetimes of the state in the bias window and the corresponding photon 
replica states get shorter and in turn less charge resides in these states.
The corresponding current transport presented in \fig{fig03}(c) demonstrates that  
only the first-excited state located just above the bias window, $1^{\rm st}$ (blue circles), is active in the current transport. 
It can be verified that the photon replica states do not participate to the current transport 
at the low electron-photon coupling strength, $g_{\gamma} = 0.001$~meV, even though $1\gamma$0 is located 
in the bias window.
The reason for the current enhancement through the $1^{\rm st}$ state is the depopulation of the 
photon replicas by the relatively large $\kappa$.

Increasing the electron-photon coupling strength to $g_{\gamma} = 0.1$~meV as shown in \fig{fig03}(b) and (d),
the occupation of the first- (blue circles) and the second-excited (red circles) states is slightly increased with $\kappa$ 
while the occupation of the ground-state and its photon replicas decreases as is demonstrated in \fig{fig03}(b).
This indicates that the lifetimes of the $1^{\rm st}$ and the $2^{\rm nd}$ get relatively longer than the lifetimes 
of the photon replica of the ground-state in the bias window. 

The states with short lifetime, and the ground-state, become active in the current transport
at higher electron-photon coupling strength, $g_{\gamma} = 0.1$~meV, 
as it is shown in \fig{fig03}(d). 
The reason for the activation the photon replica states is related to the intraband transition between 
these states. The current is going from the left lead to the right lead via the one-photon replica state located 
in the bias window while an inversion of the current transport is seen via the $1^{\rm st}$ and the $2^{\rm nd}$ 
which have a relatively longer lifetime than the photon replicas and are located outside the bias window. 
This can be understood as the following: The charge is more accumulated in the $1^{\rm st}$ and the $2^{\rm nd}$
and they are closer to the Fermi energy of the left lead leading to a current being transferred from these 
two states to the left lead appearing as a negative current. 
The participation of the photon replica states in the current transport at the higher electron-photon 
coupling strength $g_{\gamma} = 0.1$~meV enhances the total current compared to the case of the  
low electron-photon coupling strength $g_{\gamma} = 0.001$~meV. 
Therefore, the total current is enhanced with $g_{\gamma}$ as is shown in \fig{fig02}.

\subsection{Strong coupling regime, $g_{\gamma} > \kappa$}

In this section, we consider the case when the electron-photon coupling strength is stronger than 
the selected range of the cavity-environment coupling strength, $g_{\gamma} > \kappa$, 
i.e.\ the strong coupling regime.

\begin{figure}[htb]
\centering
    \includegraphics[width=0.22\textwidth]{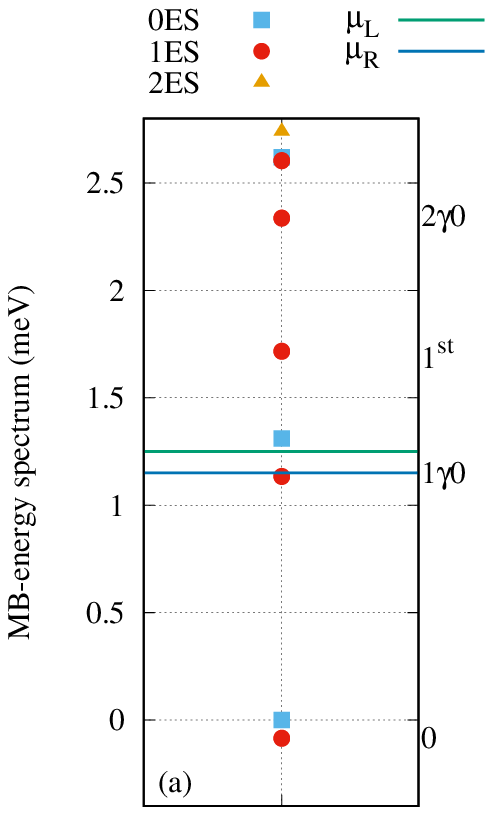}
     \includegraphics[width=0.22\textwidth]{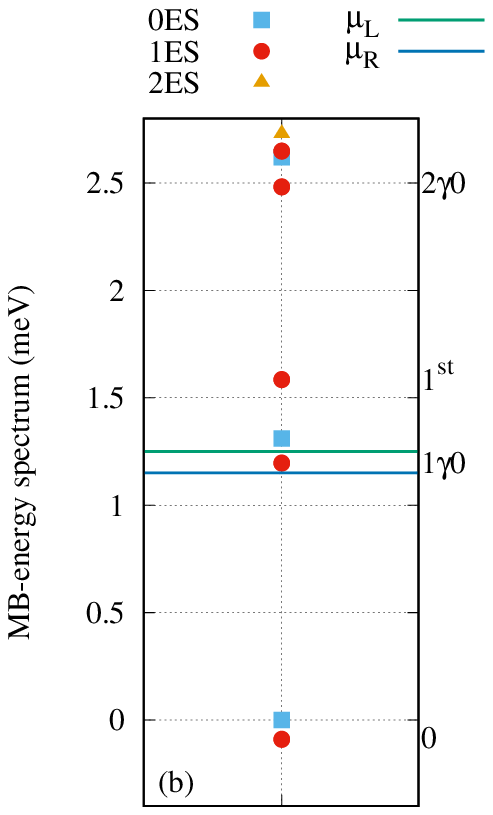}
 \caption{The many-Body energy spectra of the closed QD system coupled to the 
  photon field for the 
  $x$- (a) and $y$-polarized (b) of the photon field, where 
  0ES (blue square) are zero-electron states, 1ES (red circle) are one-electron state, 
  and 2ES (golden triangles). The photon energy is $\hbar \omega_{\gamma} = 1.31$~meV, $g_{\gamma} = 0.2$~meV, 
  and $n_\mathrm{R} = 1$. 0 represents the one-electron
  ground-state, $1^{\rm st}$ is the one-electron first-excited state,
  and 1$\gamma$0 and 2$\gamma$0 refer to the first and second photon replicas
  of the one-electron ground-state, respectively.
  The chemical potential of the left lead is $\mu_L = 1.25$~meV (green line) and the right lead is $\mu_R = 1.15$~meV (blue line).
  The magnetic field is $B = 0.1~{\rm T}$, $eV_{\rm g} = 0.651$~meV, $T_{\rm L, R} = 0.5$~K, and $\hbar \Omega_0 = 2.0~{\rm meV}$.}
\label{fig04}
\end{figure}

Figure \ref{fig04} displays the MB-energy spectrum of the QD system as a function of $\kappa$ for  
$x$- (a) and $y$-polarization (b) of the photon field when $g_{\gamma} = 0.2$ meV which is 
higher than the selected range of cavity-environment coupling strength, $g_{\gamma} > \kappa$. 
In general, tuning the electron-photon coupling strength in the case of the $x$-polarization shifts the energy 
of the states and could lead to resonant states \cite{Nzar-arXiv_article_2019_Photocurrent}. 
Comparing to the low electron-photon coupling strength, $g_{\gamma} = 0.001$ meV shown in 
\fig{fig01}, the 1$\gamma$0 moves out of the bias window and the $1^{\rm st}$ is shifted up for 
the $x$-polarization (\fig{fig04}(a)) while these changes in the $y$-polarization 
are not observed (\fig{fig04}(b)). If we further tune $g_{\gamma}$ to higher values, 
one notes a resonance between 1$^{\rm st}$ and 2$\gamma$0 in the $x$-direction (not shown). The resonant states 
will have major roles in the electron transport in the system. As we mentioned before, tuning $\kappa$ does 
not affect the energy states of the closed QD system.

\begin{figure}[htb]
\centering
    \includegraphics[width=0.45\textwidth,angle=0,bb=50 70 350 230]{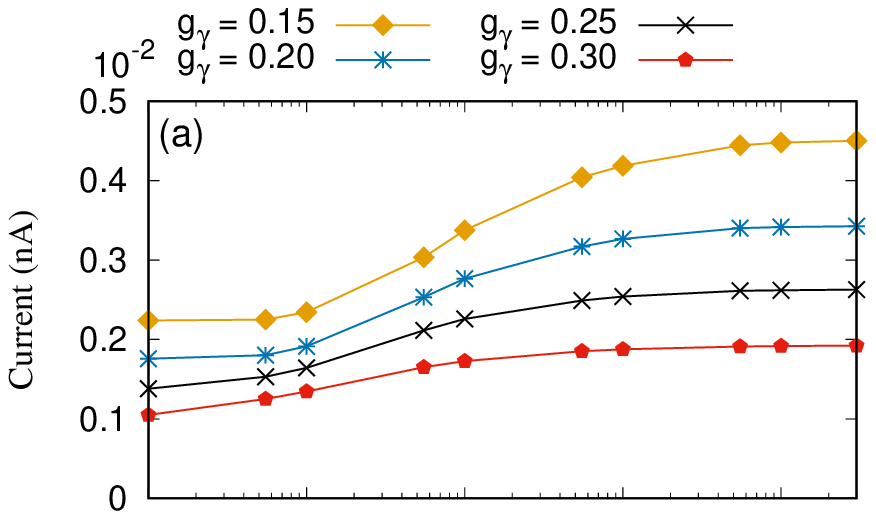}\\
    \includegraphics[width=0.45\textwidth,angle=0,bb=65 55 345 210]{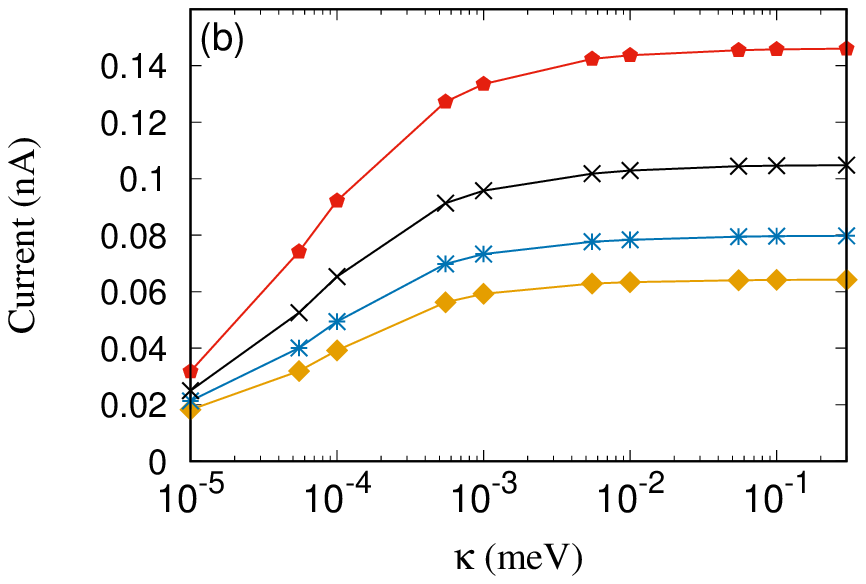}
 \caption{Current versus the the cavity-environment coupling, $\kappa$,
          for $g_{\gamma} = 0.15$ (golden diamonds), $0.2$ (blue star),  
          and $0.25$ (black cross), and $0.3$~meV (red ) in the $x$- (a) 
          and $y$-polarized of the photon field. 
          The photon energy is $\hbar\omega_{\gamma} = 1.31$~meV, and $n_\mathrm{R} = 1$.
          The chemical potential of the left lead is $\mu_L = 1.25$~meV and the right lead is $\mu_R = 1.15$~meV.
          The magnetic field is $B = 0.1~{\rm T}$, $eV_{\rm g} = 0.651$~meV, $T_{\rm L, R} = 0.5$~K, and $\hbar \Omega_0 = 2.0~{\rm meV}$}
\label{fig05}
\end{figure}

Figure \ref{fig05} displays the current versus $\kappa$ in the $x$- (a) and $y$-polarization (b) 
for different values of the electron-photon coupling strength $g_{\gamma} = 0.15$ (golden diamonds), 
$0.2$ (blue star), $0.25$ (black cross), and $0.3$~meV (red), which are all greater than the selected range of $\kappa$.
The characteristics of the current in the system is drastically changed for the $x$-polarization in
the strong coupling regime. The current is suppressed with increasing $g_{\gamma}$ for the $x$-polarization 
(\fig{fig05}(a)) throughout all values of $\kappa$ while it is enhanced in the $y$-polarization (\fig{fig05}(b)).
It indicates that the photon field with $x$-polarization is dominant at high or strong 
electron-photon coupling strength and the cavity-environment coupling effect is diminished.
In addition, the current increases with $\kappa$ in both directions of the photon polarization 
which displays a manifestation of the Purcell effect in the system, even when $g_{\gamma} > \kappa$. 
We should keep in mind that the current was enhanced with $g_\gamma$ for 
both directions of the photon polarization at the low coupling regime shown in \fig{fig02}. 

\begin{figure}[htb]
\centering
    \includegraphics[width=0.23\textwidth]{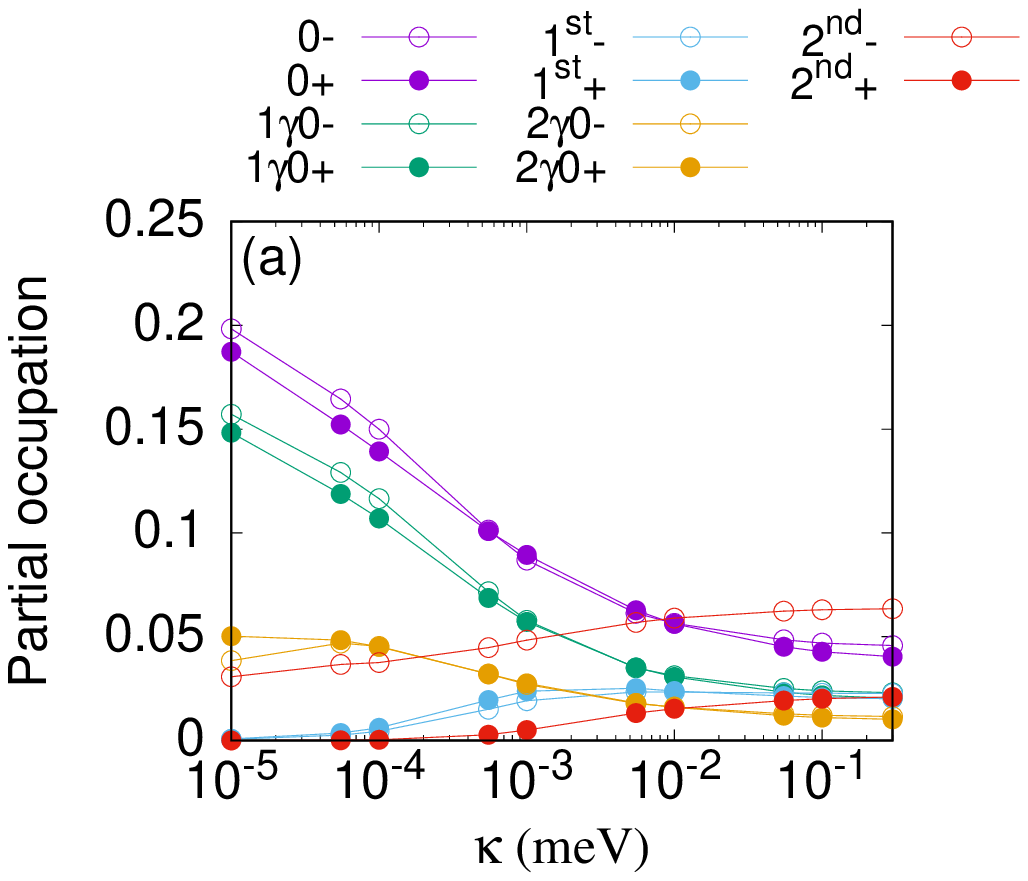}
    \includegraphics[width=0.23\textwidth]{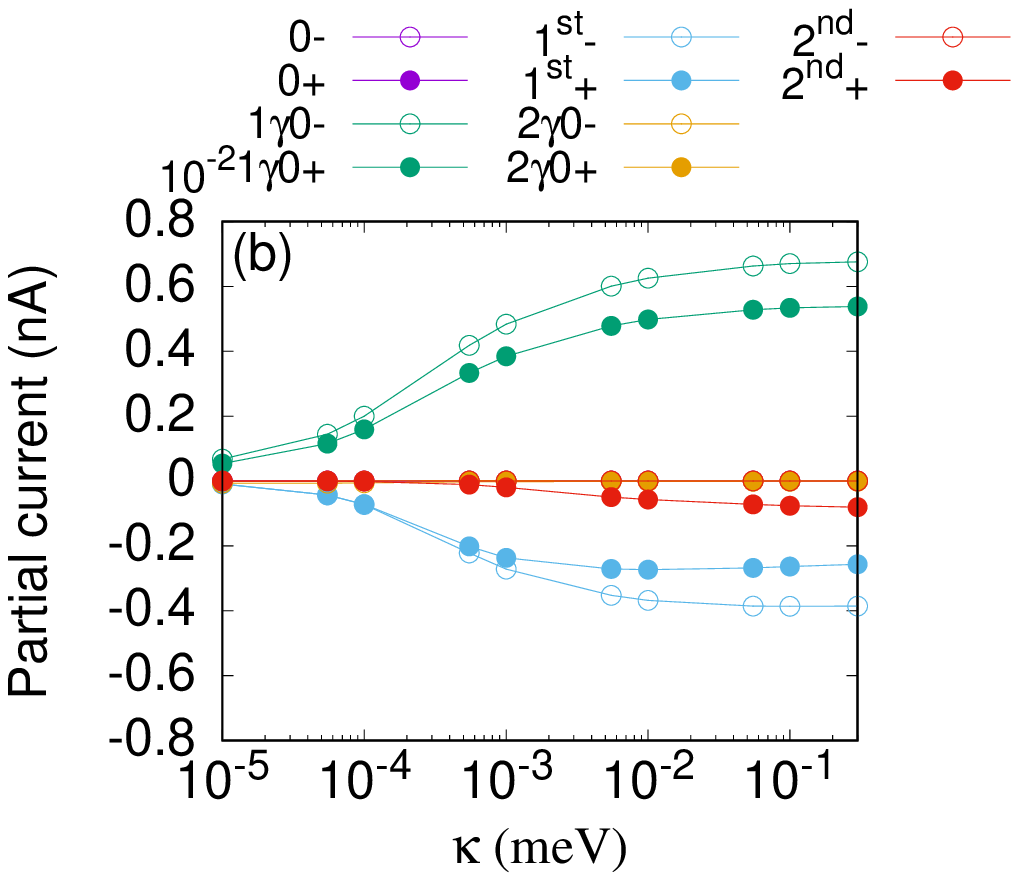}\\
 \caption{Partial occupation (a,b) and the partial current (c,d) of the ten 
          lowest one-electron states versus $\kappa$ are plotted 
          for $x$-polarization (top panel) and $y$-polarization (lower panel) when $g_{\gamma} = 0.3$~meV.
          The minus (-) and plus (+) signs indicate the Zeeman spin-down and up, respectively. 
          Herein, 0 displays the one-electron ground-state (purple circles) 
          $1\gamma$0 refers to the one-photon replica of the $0$  (green circles),
          $1^{\rm st}$ demonstrates the one-electron first-excited state (blue circles), 
          $2\gamma$0 is the two-photon replica of $0$ (golden circles),  
          and $2^{\rm nd}$ is the one-electron second-excited state (red circles).
          The photon energy is $\hbar\omega_{\gamma} = 1.31$~meV, and $n_\mathrm{R} = 1$.
          The chemical potential of the left lead is $\mu_L = 1.25$~meV and the right lead is $\mu_R = 1.15$~meV.
          The magnetic field is $B = 0.1~{\rm T}$, $eV_{\rm g} = 0.651$~meV, $T_{\rm L, R} = 0.5$~K, and
          $\hbar \Omega_0 = 2.0~{\rm meV}$.}
\label{fig06}

\end{figure}

The current suppression is related to a shift of the energy of the states that occurs at the strong coupling 
regime in the $x$-polarized photon field presented in \fig{fig04}(a). The energy shift affects the characteristics 
of the electron transport through the individual states of the QD system.
To see these effects we present the partial occupation (a) and the partial current (b) for the $x$-polarization in \fig{fig06}. 

We start with the $x$-polarized photon field with $g_{\gamma} = 0.3$~meV (\fig{fig06}(a)), 
in which the occupation of the pure electronic states, 1$^{\rm st}$ (blue) and 2$^{\rm nd}$ (red), is much enhanced 
throughout the values of $\kappa$ comparing to the case of low coupling regime when $g_{\gamma} = 0.1$~meV shown in 
\fig{fig03}(b). It supports the view that the lifetime of the pure electronic states, 1$^{\rm st}$ and 2$^{\rm nd}$, 
is much longer in the strong coupling regime increasing the accumulation of charge in these two states. 
In addition, the shift of 1$\gamma$0 outside of the bias window leads to an inactivation of the second photon 
replica state, 2$\gamma$0 to the current transport as is shown in \fig{fig06}(b). 
Therefore, the total current of the system is suppressed with increasing $g_{\gamma}$ for the $x$-polarization.
But for the $y$-polarized photon field, there is no shift of the energy of the states and thus 
the photon replica states regularly participate to the current transport leading to the current enhancement.

\section{Conclusion}\label{Sec:Conclusion}

The interplay between the electron-photon coupling and the cavity-environment coupling strengths 
on the electron transport through a quantum dot embedded in a quantum wire and coupled to 
two electron reservoirs was investigated. 
The influence of the photon polarization in the cavity on the transport properties has also been studied. 
The full electron-electron and the electron-photon interactions are taken into account using 
an exact diagonalization technique in truncated Fock many-body spaces. 
We have shown that the current is enhanced with the cavity-environment 
coupling strength for both directions of the linear photon polarization that is perpendicular or 
parallel to the electron 
transport motion in the quantum system. This is a manifestation of the Purcell effect in the system. 
In addition, the current is not linearly increased with the electron-photon coupling which is 
due to the contributing ratio of the photon-electron dressed states and the photon replica states 
to the electron transport. The contributing ratio of these states can be controlled by the electron-photon 
coupling strength.

\begin{acknowledgments}
This work was financially supported by the Research Fund of the University of Iceland,
the Icelandic Research Fund, grant no.\ 163082-051, 
and the Icelandic Infrastructure Fund. 
The computations were performed on resources provided by the Icelandic 
High Performance Computing Center at the University of Iceland.
NRA acknowledges support from University of Sulaimani and 
Komar University of Science and Technology. CST acknowledges the Ministry of
Science and Technology, Taiwan through Contract no. MOST 106-2112-M-239-001-MY3
\end{acknowledgments}

\end{document}